# Modeling Heterogeneous Melting in Phase Change Memory Devices

J. Scoggin,[1] Z. Woods,[1] H. Silva,[1] and A. Gokirmak[1]

[1]*Department of Electrical and Computer Engineering, University of Connecticut, Storrs, Connecticut 06269, USA*

We present thermodynamic crystallization and melting models and calculate phase change velocities in $Ge_2Sb_2Te_5$ based on kinetic and thermodynamic parameters. The calculated phase change velocities are strong functions of grain size, with smaller grains beginning to melt at lower temperatures. Phase change velocities are continuous functions of temperature which determine crystallization and melting rates. Hence, set and reset times as well as power and peak current requirements for switching are strong functions of grain size. Grain boundary amorphization can lead to a sufficient increase in cell resistance for small-grain phase change materials even if the whole active region does not completely amorphize. Isolated grains left in the amorphous regions, the quenched-in nuclei, facilitate templated crystal growth and significantly reduce set times for phase change memory cells. We demonstrate the significance of heterogeneous melting through 2-D electrothermal simulations coupled with a dynamic materials phase change model. Our results show reset and set times on the order of ~1 ns for 30 nm wide confined nanocrystalline (7.5 nm – 25 nm radius crystals) phase change memory cells.

Solids tend to melt heterogeneously: the liquid phase initially forms at high energy sites such as grain boundaries and material interfaces. While many materials heat ~20% above their melting temperature ($T_{melt}$) before the liquid phase forms within the bulk solid, heterogeneous melting may occur below $T_{melt}$[1]. In this manuscript, we consider the impacts of heterogeneous melting on phase change memory (PCM). PCM is a non-volatile memory technology which stores information as the low resistivity crystalline or high resistivity amorphous phase of a material (Fig. 1). PCM retention, endurance, and speed depend on the physics underlying crystallization and melting. We model temperature and grain size dependent phase change velocities in $Ge_2Sb_2Te_5$ (GST), a common phase change material, based on kinetic and thermodynamic parameters. We incorporate heterogeneous melting into a finite element phase change model coupled with electrothermal physics[2–7] and show that it can account for the experimentally demonstrated PCM performance improvement with decreasing grain size[8,9].

$T_{melt}$ is the temperature at which the Gibbs free energy difference between bulk liquid and crystalline phases ($\Delta g_{lc}$) is zero. However, melting becomes thermodynamically favorable below $T_{melt}$ at crystal interfaces. The Gibbs free energy of a spherical crystal surrounded by liquid ($\Delta G_{crys}$) is calculated by classical nucleation theory as

$$\Delta G_{crys} = -\frac{4}{3}\pi r^3 \Delta g_{lc} + 4\pi r^2 \gamma_{lc}, \quad (1)$$

where $r$ is the crystal radius and $\gamma_{lc}$ is the energy penalty at a liquid-crystal interface (Fig. 2a). (1) has extrema at $r = 0$ and $r = r_c$, the critical radius:

$$r_c = \frac{2\gamma_{lc}}{\Delta g_{lc}}. \quad (2)$$

Crystals with $r < r_c$ are subcritical and can reduce $\Delta G_{crys}$ by shrinking, i.e. melting. $r_c$ increases with $T$: $\Delta g_{lc}$ decreases with increasing $T$, crossing 0 at $T_{melt}$. $\gamma_{lc}$ is difficult to measure in GST and often used as a fitting parameter; however, $\gamma_{lc}$ increases with $T$ in metals as well as in the semiconductors Si and Ge[10–14]. Here, we use $\gamma_{lc}$ = 75 mJ/m$^2$, a temperature-independent value which allows classical nucleation theory to accurately model nucleation and growth in GST over a wide temperature range[15]. 1 nm and 10 nm radius GST grains become subcritical at ~640 K and ~840 K, respectively, with the parameters in this work (Fig. 2b).

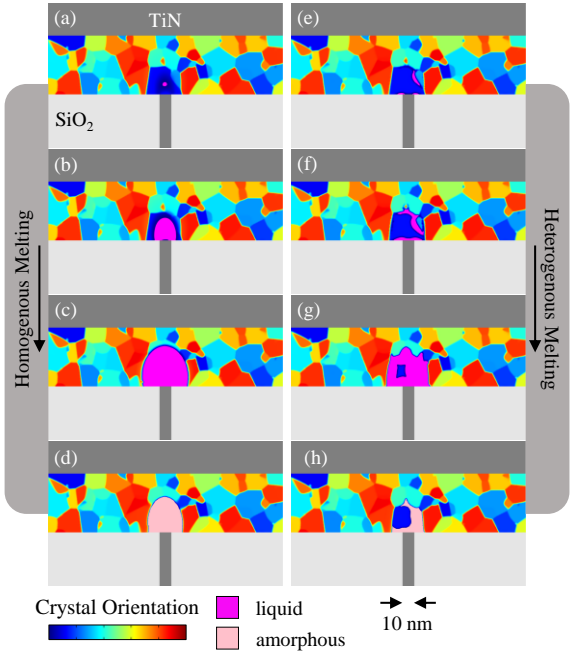

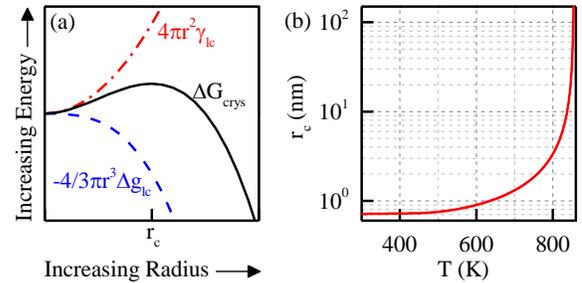

**Fig. 1**: The crystal-to-amorphous transition in a PCM mushroom cell using (a-d) homogenous and (e-h) heterogeneous melting resulting in a reset-to-initial resistance ratio of ~200 and ~100, respectively.

**Fig. 2**: (a) General behavior of volume (dashes), surface area (dash-dots), and total (solid) contributions to the Gibbs free energy of a crystal. (b) $r_c$ increases with $T$, diverging to infinity at $T_{melt}$.

Grain boundary melting may be thermodynamically favorable below $T_{melt}$ (pre-melting) even for supercritical grains ($r > r_c$) if

$$2\gamma_{lc} + w\Delta g_{lc} < \gamma_{cc}, \quad (3)$$

where $w$ is the width of liquid formed and $\gamma_{cc}$ is the crystalline-crystalline interface (grain boundary) energy penalty. Not all grain boundaries pre-melt at the same temperature since $\gamma_{cc}$ depends on the misorientation between grains, as calculated via phase field crystal simulations[16] and shown experimentally with colloidal crystals[17]. While (3) presents a thermodynamic pathway for only a finite $w$, a supercritical grain may become subcritical while pre-melting and consequently melt entirely.

Regardless of the underlying mechanism, melting is a transient process and requires kinetic and thermodynamic parameters to model. We model phase change velocity ($v$) as:

$$v = v_{kinetic}\left(1 - \exp\left(-\frac{\Delta g}{RT}\right)\right) \quad (4)$$

where $v_{kinetic}$ is the kinetic upper limit, $R$ is the gas constant, and $\Delta g$ is the non-negative thermodynamic driving force. (4) is appropriate for atomically rough interfaces[18], predicts a smooth derivative of $v$ as crystallization transitions to melt at $\Delta g = 0$ [18], and has been used to model crystallization dynamics in glass formers including GST[19,20].

We model $v_{kinetic}$ for GST as in Orava *et. al*[20]: the temperature dependence is determined from ultra-fast digital scanning calorimetry[20], and $v_{kinetic}$(900 K) is calculated from the liquid viscosity given by molecular dynamics simulations ($\eta$(900 K) ≈ 1.1 mPa s)[21].

We calculate $\Delta g$ from $\Delta g_{lc}$, which is thermodynamically related to the differences in enthalpy ($\Delta h_{lc}$) and entropy ($\Delta s_{lc}$) between phases:

$$\Delta g_{lc}(T) = \Delta h_{lc}(T) - T\Delta s_{lc}(T). \quad (5)$$

$\Delta h_{lc}$ and $\Delta s_{lc}$ can be calculated from the difference in specific heat between phases ($\Delta c_{p,lc}$):

$$\Delta h_{lc}(T) = \int_0^T \Delta c_{p,lc}(T')dT' \quad (6a)$$

$$= \Delta h_{lc}(T_x) + \int_{T_x}^T \Delta c_{p,lc}(T')dT' \quad (6b)$$

$$\Delta s_{lc}(T) = \int_0^T \frac{\Delta c_{p,lc}(T')}{T'}dT' \quad (7a)$$

$$= \Delta s_{lc}(T_y) + \int_{T_y}^T \frac{\Delta c_{p,lc}(T')}{T'}dT' \quad (7b)$$

where $T_x$ and $T_y$ are temperatures at which $\Delta h_{lc}$ and $\Delta s_{lc}$ are known. We treat the amorphous and liquid phases as a single material with a sharp change in thermodynamic parameters at the glass transition temperature ($T_{glass}$ = 431 K[22]), neglecting the dependence of these parameters and $T_{glass}$ on thermal history[23]. We model liquid and crystalline specific heats such that $\Delta c_{p,lc}$ accounts for the difference in the latent heats of fusion and crystallization[9] (Fig. 3a). We use $\Delta h_{lc}$(916 K) = 128.9 J/g[22] and $\Delta s_{lc}(T_{melt}) = \Delta h_{lc}(T_{melt})/T_{melt}$ as our known values ($T_{melt}$ = 855 K[24]) and calculate $\Delta h_{lc}(T)$, $\Delta s_{lc}(T)$, and $\Delta g_{lc}(T)$ (Fig. 3b). We calculate phase change velocities using $\Delta g_{lc}$ for bulk ($v_\infty$) and $\Delta g_{crys}$ for crystals with radius $r$ ($v_r$):

$$\Delta g_{crys} = \Delta G_{crys} \times \frac{m_{GST}}{Vol_r \times d_{GST}} \quad (8)$$

where $m$ is molar mass, $Vol$ is volume, and $d$ is mass density. We also use $m_{GST}$ and $d_{GST}$ 8to convert $\Delta g_{lc}$ from J/g to J/mol before using it in (4). $v$ is a continuous function which changes sign at the size dependent critical temperature (i.e. when $r = r_c$, Fig. 2b), implemented using $\Delta g_{cl} = -\Delta g_{lc}$ (Fig. 3c). Smaller grains are expected to melt faster or crystallize slower. Hence using $v_\infty$, as in the simulations below, gives a lower bound for melting and an upper bound for crystallization velocities.

We model heat transfer and current continuity to simulate PCM device operation including thermoelectric effects[26]:

$$dc_p\frac{dT}{dt} - \nabla \cdot (k\nabla T) = -\nabla V \cdot J - \nabla \cdot (JST) + q_H \quad (9)$$

$$\nabla \cdot J = \nabla \cdot (-\sigma\nabla V - \sigma S\nabla T) = 0 \quad (10)$$

where $k$ is thermal conductivity, $V$ is electric potential, $J$ is current density, $S$ is the Seebeck coefficient, $q_H$ is the latent heat of phase change[27], and $\sigma$ is electrical conductivity.

We model phase change similarly as in Woods *et. al*[2,3], which we briefly describe here before discussing updates that capture heterogeneous melting. Woods *et. al*[2,3] tracks the 3-vector $\overrightarrow{CD}$ with coupled rate equations

$$\frac{dCD_i}{dt} = Nucleation_i + Growth_i + Melt_i. \quad (11)$$

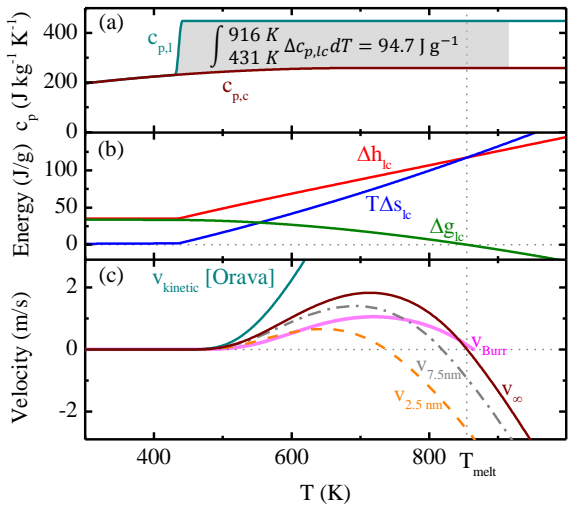

**Fig. 3**: (a) The integral of $\Delta c_{p,lc}$ accounts for the difference in the latent heats of fusion and crystallization ($\Delta h_{lc}$(916 K) = 128.9 J/g and $\Delta h_{lc}$(431 K) = 34.2 J/g, respectively[22]). (b) Calculated thermodynamic parameters. (c) Phase change velocities for bulk, 7.5 nm (dash-dots), and 2.5 nm (dashes) radius grains and that calculated by Burr *et. al*[25]. Negative values denote melting. $v_{kinetic}$ stabilizes with the liquid viscosity at higher temperatures; $v_{kinetic}(T > 2000$ K$)$ ≈ 60 m/s.

The phase of the material (*CD*) is defined by the sum of components: $CD = \Sigma(CD_i) = 0$ for the amorphous/liquid or 1 for the crystalline phase. Grain orientation is given by the distribution of $CD_i$ values (Fig 4a), with grain boundaries defined where $|\nabla \overrightarrow{CD}| > 5\times10^{-3}$ nm$^{-1}$. *Nucleation*$_i$ randomly generates nuclei at a temperature dependent rate. *Growth*$_i$ has a stability term (*stbl*$_i$) which drives *CD* to 0 or 1 and a diffusivity term (*diff*$_i$) which grows grains outwards:

$$Growth_i = stbl_i \times \frac{CD_i}{CD} + \nabla(diff_i)\nabla CD_i$$
$$stbl_i = \alpha_{stbl} \times v(T) \times pw_{stbl}(CD) \quad (12)$$
$$diff_i = \alpha_{diff} \times v(T) \times pw_{diff}(CD)$$

where $\alpha_{stbl} = 0.8$ nm$^{-1}$ and $\alpha_{diff} = -0.2$ nm are constants, $pw_{stbl}$ and $pw_{diff}$ are piecewise control functions (Fig 5), and *v* is the phase change velocity. A valley in $pw_{stbl}$ prevents small perturbations above 0 from triggering crystallization (Fig 5a inset). Nucleation or templated growth from an adjacent crystal is required to escape this well. *Melt*$_i$ brings $CD_i$ (and thus *CD*) to 0 when $T > T_{melt}$:

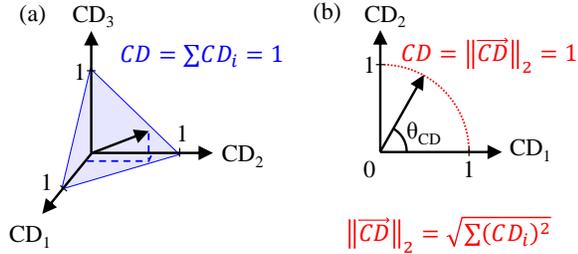

**Fig 4.** Crystalline vectors defined by (a) a 3-vector with $CD = \Sigma(CD_i)$ as in Woods *et. al*[3] and (b) a 2-vector with $CD = \|CD\|_2$ as used here. The cut plane and dotted line show allowed crystalline values.

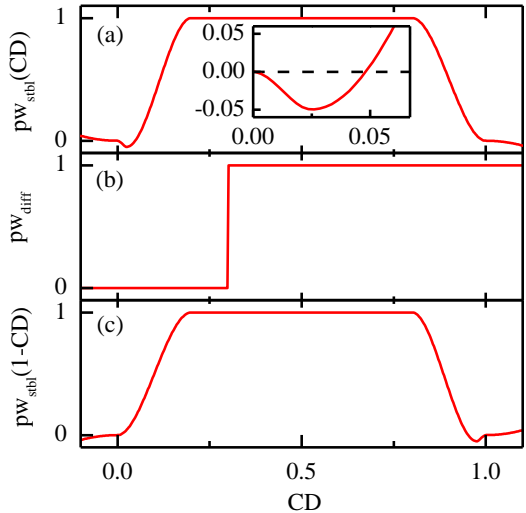

**Fig 5**. Control functions for (a) *stbl*$_i$ when $T < T_{melt}$, (b) *diff*$_i$, and (c) *stbl*$_i$ when $T > T_{melt}$. A stability well in (a)/(c) prevents small perturbations from triggering crystallization/melt, respectively.

$$Melt_i = \alpha_{melt} \times (T > T_{melt}) \times CD_i \quad (13)$$

where $\alpha_{melt} = -1$ ns$^{-1}$ is a constant and $(T > T_{melt})$ is a step function (0 to 1 over a 1 K window centered at $T_{melt}$).

In this work, we use a 2-vector $\overrightarrow{CD}$, which is sufficient to capture grain boundaries while reducing the number of equations solved. We use $CD = \sqrt{CD_1^2 + CD_2^2}$ instead of $CD = \Sigma(CD_i)$, fixing all crystalline vectors at an equal length in *CD*-space and ensuring that their lengths approach zero (they melt) at the same rate (Fig 4b). *Nucleation*$_i$ generates nuclei with a random angle ($10° \leq \theta_{CD} \leq 80°$). $\theta_{CD}$ can be mapped to a physical grain orientation ($\theta_{phys}$) with a function that depends on the dimensionality and crystalline structure of interest, e.g. to a range of $0° \leq \theta_{phys} < 90°$ for 2-D simple cubic structures or $0° \leq \theta_{phys} < 60°$ for 2-D hexagonal structures. We define grain boundaries wherever there is a high gradient in $\theta_{CD}$ ($|\nabla\theta_{CD}| > 2.9$ °/nm).

We update $v(T)$ from the velocity curve given by Burr *et. al*[25] to $v_\infty(T)$. The two curves are qualitatively similar for $T < T_{melt}$, but $v_\infty(T)$ has a higher peak velocity and is also defined for $T > T_{melt}$ (Fig 3c). We initiate melting at grain boundaries and material interfaces by modifying *Melt*$_i$:

$$Melt_i = \alpha_{melt} \times v(T) \times (T > T_{melt}) \times CD_i \times HM, (14)$$

where *HM* is 1 at heterogeneous melting sites (grain boundaries and material interfaces) and 0 elsewhere. We modify *stbl*$_i$ to have a stability valley at $CD \lessapprox 1$ when $T > T_{melt}$ to maintain well-defined grains during melt (Fig 5c):

$$stbl_i = \text{sign}(CD_i) \times \alpha_{stbl} \times v(T)$$
$$\times \begin{pmatrix} (T < T_{melt}) \times pw_{stbl}(\text{sign}(CD_i) \times CD) \\ +(T > T_{melt}) \times pw_{stbl}(1 - \text{sign}(CD_i) \times CD) \end{pmatrix}. (15)$$

We use the *sign* function (-1/+1/0 for negative/positive/0 arguments, respectively) in (15) to properly call control functions with the non-negative $CD = \sqrt{CD_1^2 + CD_2^2}$ if $CD_i$ becomes negative due to numerical errors. We can control the temperature at which melting begins ($T_{melt}$), the time required for a liquid layer to form between grains (via $\alpha_{melt}$), and the melting rate [$v(T)$, $T > T_{melt}$] with this framework.

To demonstrate this updated model, we first simulate heterogeneous melting of a polycrystalline 2-D 200 nm square (20 nm out of plane depth) by setting $T = 900$ K (Fig 6). A liquid layer forms at grain boundaries and around the perimeter in ~1 ns, then grows at $v_\infty(900\,K) \approx 1.3$ nm/ns. Next, we simulate electrical cycling of a 30 nm wide confined cell (30 nm out of plane depth). We use a regular array of equally sized grains for the initial crystallinity conditions rather than a random grain map so that we can explicitly define grain size (Fig 7a). We vary the durations of square current pulses with 0.1 ns rise and fall times for reset and set of devices with 7.5 nm, 15 nm, and 25 nm radius grains. The devices achieve a reset resistance increase ($R/R_{Initial}$) of $>10^2$ in 4.2 ns when $r = 7.5$ nm but require 5.2 ns to reach this contrast when $r = 15$ nm or 25 nm (Fig 7b, Fig 8). We use devices reset to similar $R/R_{Initial}$ for set comparisons. 7.5 nm grains set more quickly due to templated growth from quenched-in crystals when $R/R_{Initial} \approx 20$ (Fig 7c, Fig 9). However, fewer quenched-in crystals remain after stronger resets, increasing the resistance ratio and set times: set is achieved through templated growth from crystalline fronts at the top and bottom contacts. Wang *et. al*[8] experimentally showed a trend of decreased reset and set times as radii decreased from 8.5 nm to 5 nm but achieved sub-nanosecond reset while requiring ~50 ns for set. The longer reset and shorter set times in our simulations are consistent with a $v_\infty$ that underestimates melting velocities but overestimates crystallization velocities, resulting in less melting during reset, more crystallization while cooling after reset, and faster crystallization during set.

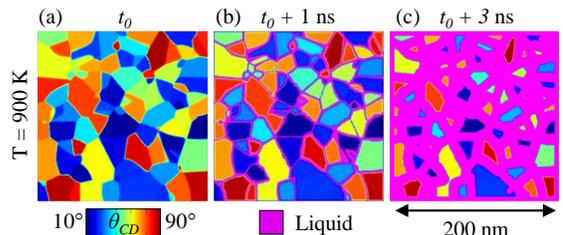

**Fig 6**. (a) A polycrystalline GST square is brought to 900 K at $t_0$. (b) Liquid forms at grain boundaries in 1 ns and (c) melts at 1.3 nm/ns.

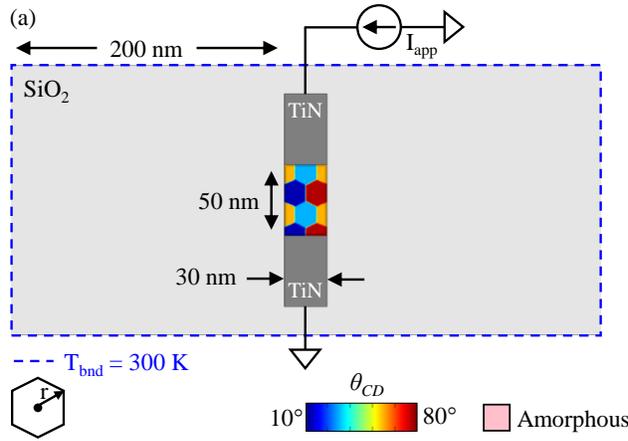
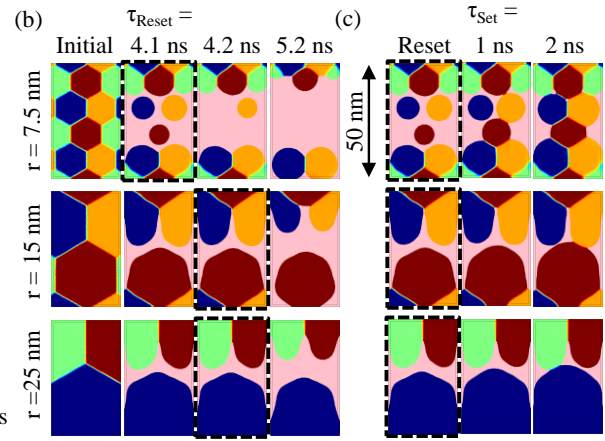

**Fig 7.** (a) Schematic illustration of geometry and initial and boundary conditions used in simulations. $I_{app}$ is a square pulse with 0.1 ns rise and fall times and a magnitude of 500 µA for reset or 50 µA for set. (b) Crystal maps of initial conditions and after reset with 3 different pulse durations ($\tau_{Reset}$) for 7.5 nm, 15 nm, and 25 nm grains. The amorphous area (and hence reset resistance) increases with $\tau_{Reset}$. The *Initial* column is used for $R_{Initial}$ in Fig 8 and Fig 9. A ~20x increase in cell resistance requires a 4.1 ns reset pulse for 7.5 nm grains and a 4.2 ns reset pulse for 15 nm or 25 nm grains (dashed borders). (c) Quenched-in crystals embedded in the amorphous GST in the $r = 7.5$ nm case grow simultaneously, allowing for shorter set times.

In conclusion, we have proposed and studied heterogeneous melting as a mechanism for improved PCM performance as grain size decreases. Smaller grains result in more phase change sites acting simultaneously, decreasing reset and set times. The models presented are for $Ge_2Sb_2Te_5$, but temperature and grain size dependent phase change velocities for other materials can be calculated given the appropriate thermodynamic ($c_{p,l}$, $c_{p,c}$, $\Delta h_{lc}(T_x)$, and $T_{melt}$) and kinetic [$\eta(T)$] parameters.

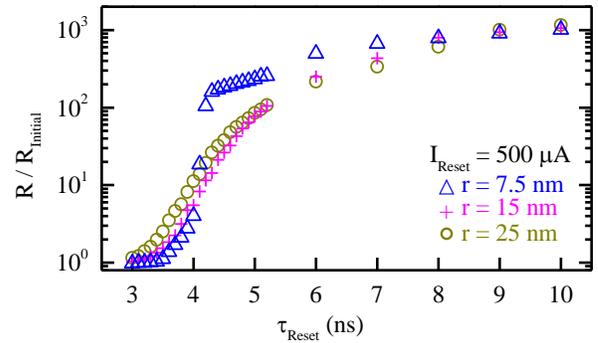

**Fig 8.** 'Reset' resistance after a square current pulse (500 µA magnitude and $\tau_{Reset}$ duration). Smaller grains allow for faster reset, while larger grains have a wider pulse duration window for intermediate values. $R_{Initial}$ corresponds to the cells in the *Initial* column in Fig 7b.

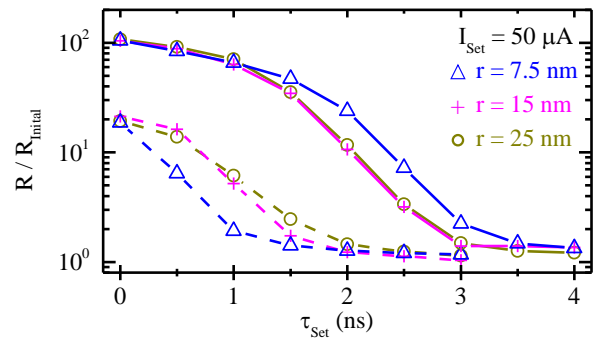

**Fig 9.** 'Set' resistance after a square current pulse (50 µA magnitude and $\tau_{Set}$ duration) beginning with ~20x (dashed lines) and ~100x (solid lines) reset resistances for each grain size (Fig 8). Smaller grains allow for faster set when the initial reset resistance is low. Set time becomes less dependent on grain size as fewer quenched-in crystals remain in the amorphous region and reset resistance increases; crystallization is achieved through templated growth from the same number of similarly spaced crystal-amorphous interfaces. $R_{Initial}$ corresponds to the cells in the *Initial* column in Fig 7b.